\documentstyle[psfig,eqsecnum,aps]{revtex}
\begin{document}
\title{Beta decay in odd-A and even-even proton-rich Kr isotopes}
\author{P. Sarriguren, E. Moya de Guerra, and A. Escuderos}
\address{Instituto de Estructura de la Materia,\\
Consejo Superior de Investigaciones Cient\'{\i }ficas, \\
Serrano 123, E-28006 Madrid, Spain}
\maketitle

\begin{abstract}
$\beta$-decay properties of proton-rich odd-A and even-even Krypton 
isotopes are studied in the framework of a deformed selfconsistent 
Hartree-Fock calculation with density-dependent Skyrme forces,  
including pairing correlations between like nucleons in BCS approximation. 
Residual spin-isospin interactions are consistently included in the 
particle-hole and particle-particle channels and treated in Quasiparticle 
Random Phase Approximation. The similarities and differences in the 
treatment of even-even and odd-A nuclei are stressed. Comparison to
available experimental information  is done for Gamow-Teller strength
distributions, summed strengths, and half-lives. The dependence of these
observables on deformation is particularly emphasized in a search for
signatures of the shape of the parent nucleus.

\end{abstract}


\section{Introduction}

One of the most exciting challenges in current Nuclear Physics is the 
understanding of nuclear systems under extreme conditions \cite{review}. 
The opportunities offered by recent experimental work using beams 
of exotic nuclei and the corresponding theoretical efforts to describe them 
are of great interest, especially for nuclear structure physics and nuclear 
astrophysics \cite{astro}.

The decay properties and cross sections for nuclear reactions of radioactive 
nuclei are fundamental to understand various phases in the stellar evolution,
including the energy generation, the nucleosynthesis, and the abundance of 
elements. Since this information cannot be determined
experimentally for the extreme conditions of temperature and density that
hold in the interior of the star, reliable theoretical calculations for
these processes are mandatory.
In particular, the decay properties of proton rich nuclei are fundamental 
to understand the $rp$ process (rapid proton capture nucleosynthesis),
characterized by very large proton capture reaction rates on proton rich 
nuclei \cite{schatz}. Of special importance in this context are the waiting 
points like $^{72}$Kr, where the $rp$ process is inhibited and the reaction 
flow has to wait for the relatively slow $\beta$-decay to continue. The total
half-lives of the waiting points determine the speed of nucleosynthesis towards
heavier nuclei as well as the isotopic abundances.

Decay properties and nuclear structure are intimately related.
It is clear that a precise and  reliable description of the ground state of the
parent nucleus and of the states populated in the daughter nucleus is
necessary to obtain a good description of the decay, and vice versa, 
failures to describe the decay properties would
indicate that an improvement of the theoretical formalism is needed.

From a theoretical point of view the physics of exotic nuclei, characterized 
by very unusual ratios of neutrons and protons, can be considered as a test 
for the already well established models of nuclear structure that are used
to describe stable systems. Since the parameters and interactions used in 
the usual shell model or mean field calculations are determined in order 
to reproduce the properties of known nuclei, they may not always 
be appropriate for use in the calculations of nuclei approaching the drip lines.

Different microscopic models to describe the $\beta $-strength are present in 
the literature. Models based on spherical single-particle wave functions and 
energies with pairing and residual interactions treated in Quasiparticle
Random Phase Approximation (QRPA) were first studied in Ref. \cite{sph}. 
Deformation was included in Ref. \cite{kru}, where a Nilsson potential was
used to generate single-particle orbitals. Extensions including Woods-Saxon 
type potentials \cite{moll}, residual interactions in the particle-particle 
channel \cite{hir}, Hartree-Fock (HF) mean field with separable residual 
interactions treated in Tamm Dancoff approximation \cite{frisk}, selfconsistent 
approaches in spherical neutron-rich nuclei \cite{doba} and based on an 
energy-density functional \cite{borzov}, can be also found in the literature.

In a previous work \cite{sarr1,sarr2,sarr4} we studied ground state and 
$\beta $-decay properties of even-even exotic nuclei on the basis of a deformed 
selfconsistent HF+BCS+QRPA calculation with density dependent effective 
interactions of Skyrme type.
This is a well founded approach that has been very successful in the 
description of spherical and deformed nuclei within the valley of 
stability \cite{flocard}. In this method once the parameters of the effective 
Skyrme interaction are determined, basically  by fits
to global properties in spherical nuclei over the nuclear chart, and the gap
parameters of the usual pairing force and the coupling strength of the residual
neutron-proton pairing force are specified, there are no free parameters
left. Both the residual interaction and the mean field are consistently obtained
from the same two-body force. 
This is therefore a reliable method, suitable for extrapolations into the 
unstable regions approaching the drip lines. It is worth investigating
whether these powerful tools designed to account for the properties of
stable nuclei are still valid when approaching the drip lines.

One possible way to establish the validity of the known approaches as well as 
the limits of their applicability, is a systematic investigation of 
nuclei covering the whole range from stability towards the drip lines.
Exploration of series of isotopes moving away from the region of stability
would fulfill these requirements allowing us to learn how the adequacy of 
the description evolves in progressively more unstable nuclei.

Following the same criteria as in our previous work \cite{sarr1,sarr2,sarr4}, 
we apply this
formalism to the study of proton rich Krypton isotopes, including even-even as 
well as odd-A isotopes for the first time.
The reasons why this is a region of special interest to study 
$\beta $-decay have already been stressed in Refs. \cite{sarr1,sarr2,sarr4}. 
They include the large
$Q$-values in proton rich nuclei \cite{hamasag}, the competition of nuclear
shapes \cite{frisk,petro} that characterizes this mass region, and the 
possibility to approach systematically the $N=Z$ isotope. Thus, we can test 
the validity of our formalism and look for discrepancies when approaching the
drip lines.

The paper is organized as follows. We first summarize in Sect. 2 the method
of calculation. Mean field and QRPA with residual spin-isospin interactions 
are introduced and explained separately for both even-even and odd-A nuclei. 
Gamow-Teller strengths, excitation energies, $Q_{EC}$ values and half-lives
are also discussed. In Sect. 3 we present our results. We discuss similarities
and differences in even-even and odd-A nuclei and analyze in detail the 
$\beta$-decay observables in Kr-isotopes, comparing them with the available 
experimental data.
Sect. 4 contains the conclusions and some final remarks.

\section{Theoretical Formalism}

In this Section we summarize briefly the theory involved in the microscopic
calculations. More details can be found in
Refs. \cite{sarr1,sarr2,sarr4}. Our method consists in a selfconsistent
formalism based on a deformed Hartree-Fock  mean field obtained with a
Skyrme interaction, including pairing correlations in the BCS approximation.
We consider in this paper the force SG2 \cite{giai} of Van Giai and Sagawa,
that has been successfully tested against spin and isospin excitations in 
spherical \cite{giai} and deformed nuclei \cite{sarr3}.
Comparison to calculations obtained with other Skyrme forces have been made
in Refs. \cite{sarr1,sarr2}, showing that the results do not differ in a 
significant way.
The single particle energies, wave functions, and occupation probabilities 
are generated from this mean field. 

For the solution of the HF equations we follow the McMaster procedure that
is based in the formalism developed in Ref.\cite{vautherin} and described in
Ref.\cite{vallieres}. Time reversal and axial symmetry are assumed.
The single-particle wave functions are expanded in
terms of the eigenstates of an axially symmetric harmonic oscillator in
cylindrical coordinates. We use eleven major shells. The method also
includes pairing between like nucleons in the BCS approximation with fixed
gap parameters for protons $\Delta _{\pi},$ and neutrons $\Delta _{\nu}$, which
are determined phenomenologically from the odd-even mass differences through
a symmetric five term formula involving the experimental binding energies 
\cite{audi}. The values used in this work are the same as those given in Ref.
\cite{sarr2}.

For odd-A nuclei, the fields corresponding to the different interactions were
obtained by doing one iteration from the corresponding selfconsistent field
of the closest even-even nucleus, selecting the orbital occupied by the odd
nucleon according to the experimental spin and parity. For those cases where
this experimental assignment is not well established we choose the orbital
closer to the Fermi level. The chosen state is blocked  from the BCS 
calculation, and we assign to it a pair occupation probability of 0.5.
The effect of doing several more iterations from the even-even case, in order
to see how the extra particle polarizes the core, was studied in 
Ref. \cite{sprung} without observing significant changes. 
We repeated these calculations here looking for some effect on the GT strength 
distributions, but again the changes were negligible.
According to this procedure our spin and parity assignments are as follows:
$5/2^+$ for $^{77}$Kr and $^{75}$Kr (as determined experimentally \cite{audi}),
$3/2^-$ for $^{73}$Kr from the most recent experimental 
determination \cite{exp73}. 
For $^{69}$Kr and  $^{71}$Kr we take the spin and parity
according to our calculations and assign $K^{\pi}=1/2^-$ in the first case
and  $K^{\pi}=9/2^+$ or  $K^{\pi}=3/2^-$ in  $^{71}$Kr depending on the 
oblate or prolate shape.
We describe the even $Z$ odd $N$ parent nucleus by removing one neutron from the 
selfconsistent field of the even-even nucleus and the odd $Z$ even $N$ daughter 
nucleus by removing one proton. This is a proper way to describe both the parent 
and daughter nuclei from the same mean field.
Taking $(Z,N)$ as the even-even nucleus for reference, the parent 
odd-A nucleus is $(Z,N-1)$ and it decays into the daughter nucleus $(Z-1,N)$.
As an example, to describe the $\beta ^+$ decay of the odd-neutron parent nucleus
$^{73}$Kr (Z=36, N=37) into the odd-proton daughter nucleus $^{73}$Br 
$(Z=35, N=38)$, we use the mean field of the  even-even nucleus $^{74}$Kr 
$(Z=36, N=38)$.

In a previous work \cite{sarr2} we analyzed the energy surfaces as a function 
of deformation for all the even-even isotopes under study here. For that 
purpose, we performed constrained
HF calculations with a quadrupole constraint \cite{constraint} and we 
minimized the HF energy under the constraint of keeping fixed the nuclear 
deformation. Calculations in this paper are performed for the equilibrium 
shapes of each nucleus obtained in that way, that is, for the solutions, 
in general deformed, for which we obtained minima in the energy surfaces. 
Most of these nuclei present oblate and prolate equilibrium 
shapes \cite{sarr2} that are very close in energy.
As we have mentioned, single-particle energies, wave functions, and
occupation probabilities in the odd nuclei, are obtained from the
converged mean fields of the even-even neighbor. Therefore, calculations 
in the odd nuclei are 
done at the corresponding equilibrium shapes of the even-even generator.

In this work, besides the odd-nuclei, we also consider a new even-even 
isotope, $^{70}$Kr,
that was not previously included in the isotope chain studied in 
Refs. \cite{sarr2,sarr4}. We then show
first in Fig. 1 the total energy of $^{70}$Kr as a function of the mass
quadrupole moment $Q_0$. The results correspond to a constrained HF+BCS
calculation with the Skyrme forces SG2 and Sk3. As one can see, 
we obtain two minima in the energy profile, one is oblate and the other is
prolate at about the same value of the quadrupole moment. These two 
minima are separated by about 1 MeV, the oblate being the deepest one
with the two forces. Thus, $^{70}$Kr follows nicely the trend observed
for $^{72,74,76,78}$Kr isotopes in Ref. \cite{sarr2}, with an oblate shape 
favored in a shape coexistent isotope.

\subsection{Residual interactions}

To describe Gamow-Teller transitions
we add to the mean field a spin-isospin residual
interaction, which is expected to be the most relevant interaction for that 
purpose. This interaction contains two parts, particle-hole ($ph$) and 
particle-particle ($pp$). The $ph$ part is responsible 
for the position and structure of the GT resonance \cite{hir,sarr2} and
is derived selfconsistently from the same energy density functional (and Skyrme 
interaction) as the HF equation, in terms of the second derivatives of
the energy density functional with respect to the one-body densities 
\cite{bertsch}. The $ph$ residual interaction is finally written in a separable
form by averaging the Landau-Migdal resulting force over the nuclear volume,
as explained in Refs. \cite{sarr1,sarr2}.
The coupling strength $\chi ^{ph}_{GT}$ is
completely determined by the Skyrme parameters,
the nuclear radius, and the Fermi momentum.

The particle-particle part is a neutron-proton pairing force in the $J^\pi=1^+$
coupling channel.
We introduce this interaction in the usual way \cite{hir,sarr4,kpp,muto}, that is, 
in terms of a 
separable force with a coupling constant $\kappa ^{pp}_{GT}$, which is fitted 
to the phenomenology. Since the peak of the
GT resonance is almost insensitive to the $pp$ force, $\kappa ^{pp}_{GT}$ 
is usually adjusted to reproduce the half-lives \cite{hir}.
However, one should be careful with the choice of this coupling constant.
Since the $pp$ force is introduced independently of the mean field, 
if $\kappa ^{pp}_{GT}$ is strong enough it may happen that the QRPA
collapses, because the condition that the ground state be stable against the
corresponding mode is not fulfilled. This happens because the $pp$ force, being 
an attractive force, makes the 
GT strength to be pushed down to lower energies with increasing values of 
$\kappa ^{pp}_{GT}$. 
A careful search of the optimal strength can certainly be done for each 
particular case, but this is not our purpose in this work. Instead, we have chosen
the same coupling constant ($\kappa ^{pp}_{GT}=0.07$ MeV) for all nuclei 
considered here. This value was obtained \cite{sarr4} under the requirements of 
improving in general the agreement with
experimental half-lives of even-even isotopes in this mass region
while being still far from the values leading
to the collapse. 

\subsection{Even-even nuclei}

The proton-neutron QRPA phonon operator for GT excitations in
even-even nuclei is written as

\begin{equation}
\Gamma _{\omega _{K}}^{+}=\sum_{\pi\nu}\left[ X_{\pi\nu}^{\omega _{K}}\alpha
_{\nu}^{+}\alpha _{\bar{\pi}}^{+}-Y_{\pi\nu}^{\omega _{K}}\alpha _{\bar{\nu}}\alpha
_{\pi}\right]\, ,  
\label{phon}
\end{equation}
where $\alpha ^{+}\left( \alpha \right) $ are quasiparticle creation
(annihilation) operators, $\omega _{K}$ are the excitation energies, and 
$X_{\pi\nu}^{\omega _{K}},Y_{\pi\nu}^{\omega _{K}}$ the forward and backward
amplitudes, respectively. From the QRPA equations the forward and backward 
amplitudes are obtained as \cite{muto}
\begin{equation}
X_{\pi\nu}^{\omega _{K}}=\frac{1}{\omega _{K}-{\epsilon}_{\pi\nu}}
\left[ 2\chi ^{ph}_{GT} \left( q_{\pi\nu}{M}_{-}^{\omega _{K}}+\tilde{q}_{\pi\nu}
{M}_{+}^{\omega _{K}}\right)
-2\kappa ^{pp}_{GT} \left( q^{U}_{\pi\nu}{M}_{--}^{\omega _{K}}+q^{V}_{\pi\nu}
{M}_{++}^{\omega _{K}}\right) \right] \, ,
\label{xphon}
\end{equation}

\begin{equation}
Y_{\pi\nu}^{\omega _{K}}=\frac{-1}{\omega _{K}+{\epsilon}_{\pi\nu}}
\left[ 2\chi ^{ph}_{GT} \left( q_{\pi\nu}{M}_{+}^{\omega _{K}}+\tilde{q}_{\pi\nu}
{M}_{-}^{\omega _{K}}\right)
+2\kappa ^{pp}_{GT} \left( q^U_{\pi\nu}{M}_{++}^{\omega _{K}}+q^V_{\pi\nu}
{M}_{--}^{\omega _{K}}\right) \right] \, ,
\label{yphon}
\end{equation}
with ${\epsilon}_{\pi\nu}=E_{\nu}+E_{\pi}$ the two-quasiparticle excitation
energies in terms of the quasiparticle energies $E_{i}$. 
${M}^{\omega _{K}}$ are given by

\begin{equation}
M_{-}^{\omega _{K}}=\sum_{\pi\nu}\left( q_{\pi\nu}X_{\pi\nu}^{\omega _{K}}+
\tilde{q}_{\pi\nu}Y_{\pi\nu}^{\omega _{K}}\right) \, ,
\label{m-}
\end{equation}

\begin{equation}
M_{+}^{\omega _{K}}=\sum_{\pi\nu}\left( \tilde{q}_{\pi\nu}
X_{\pi\nu}^{\omega _{K}}+
q_{\pi\nu}Y_{\pi\nu}^{\omega _{K}}\right) \, ,
\label{m+}
\end{equation}

\begin{equation}
M_{--}^{\omega _{K}}=\sum_{\pi\nu}\left( q^U_{\pi\nu}X_{\pi\nu}^{\omega _{K}}-
q^V_{\pi\nu}Y_{\pi\nu}^{\omega _{K}}\right) \, ,
\label{m--}
\end{equation}%

\begin{equation}
M_{++}^{\omega _{K}}=\sum_{\pi\nu}\left( q^V_{\pi\nu}X_{\pi\nu}^{\omega _{K}}-
q^U_{\pi\nu}Y_{\pi\nu}^{\omega _{K}}\right) \, ,
\label{m++}
\end{equation}
with
\begin{equation}
\tilde{q}_{\pi\nu}=u_{\nu}v_{\pi}\Sigma _{K}^{\nu\pi };\ \ \ 
q_{\pi\nu}=v_{\nu}u_{\pi}\Sigma _{K}^{\nu\pi};\ \ \ 
q^V_{\pi\nu}=v_{\nu}v_{\pi}\Sigma _{K}^{\nu\pi};\ \ \ 
q^U_{\pi\nu}=u_{\nu}u_{\pi}\Sigma _{K}^{\nu\pi}\, ,
\label{qs}
\end{equation}
where $v'$s are occupation amplitudes ($u^2=1-v^2$) and $\Sigma _{K}^{\nu\pi}$ 
spin matrix elements connecting neutron and proton states with spin operators
\begin{equation}
\Sigma _{K}^{\nu\pi}=\left\langle \nu\left| \sigma _{K}\right| \pi\right\rangle 
\, .\label{sigma}
\end{equation}
Explicit expressions for these matrix elements
in the cylindrical harmonic oscillator basis can be found in Ref. \cite{sarr1}.
Note that the expressions for the $X$ and $Y$ amplitudes 
(\ref{xphon})-(\ref{yphon}) defined here differ from those in 
Refs. \cite{sarr1,sarr2} by the contributions  due to $pp$ residual interactions
that were not include there.

The solutions of the QRPA equations are obtained by solving first a dispersion
relation, which is of fourth order in the excitation energies $\omega$. 
Then, for each value of the energy the amplitudes $M$ are determined
by using the normalization condition of the phonon amplitudes.
The technical procedure to solve these QRPA equations is described in
detail in Ref. \cite{muto}.

For even-even nuclei the GT transition amplitudes in the intrinsic frame
connecting the QRPA ground state 
$\left| 0\right\rangle \ \ \left( \Gamma _{\omega _{K}} \left| 0\right\rangle 
=0 \right)$ to one phonon states $\left| \omega _K \right\rangle \ \ \left( 
\Gamma ^+ _{\omega _{K}} \left| 0\right\rangle = \left|
\omega _K \right\rangle \right)$, are given by
\begin{equation}
\left\langle \omega _K | \beta _K^{\pm} | 0 \right\rangle = \mp 
M^{\omega _K}_\pm \, .
\label{ampleven}
\end{equation}

The Ikeda sum rule is always fulfilled in our calculations.

\subsection{Odd-A nuclei}

The QRPA treatment described in the above subsection is formulated for the 
excitations of the ground state of an even-even nucleus.
The GT transition amplitudes connecting the ground state of an even-even 
nucleus (0qp state) to all one phonon states in the odd-odd daughter nucleus 
(2qp states) are given in the intrinsic frame by Eq. (\ref{ampleven}).
When the parent nucleus has an odd nucleon, the
ground state can be expressed as a 1qp state in which the odd nucleon
occupies the single-particle orbit of lowest energy. 
Then two types of transitions
are possible, which are represented schematically in Fig. 2.
i) The first type of transitions are  phonon
excitations in which the odd nucleon acts only as a spectator. We
call them three quasiparticle transitions (3qp). In the intrinsic frame,
the transition amplitudes in this case are basically the same as in the even-even
case  but with the blocked spectator excluded from the calculation,
\begin{equation}
\left< f \left|  \beta ^\pm _{K} \right| i\right>_{3qp}=
\left< \omega_K ,1qp \left| \beta_K ^{\pm }\right| 0, 1qp\right> =
\mp M^{\omega _K}_\pm \, .
\label{ampl3qp}
\end{equation}
ii) The other type of transitions are those involving the odd nucleon.
We call them one quasiparticle transitions (1qp).
We introduce as usual \cite{kru,moll,muto} phonon correlations to the 
quasiparticle transitions in first order perturbation,
taking into account the part of the GT interaction that contains terms linear 
in the $\eta_{\nu \pi}=\alpha^+_\nu \alpha_\pi$ operator.
The transition amplitudes for the correlated states can be found in 
Ref. \cite{muto}. Here, we give the explicit expression for the transition 
amplitude corresponding to the Kr isotopes, that is, a $\beta^+$-decay of an 
odd-neutron parent nucleus 
decaying into an odd-proton daughter nucleus.

\begin{eqnarray}
\left< f \left|  \beta ^+ _{K} \right| i\right>_{1qp}&=&
\left< \pi_{corr} \left| \beta ^+ _K \right| \nu_{corr}\right> = \nonumber \\ 
&& q^V_{\pi\nu}+2\chi ^{ph}_{GT} \left\{ q^V_{\pi\nu} \sum_{\omega_K}\left[ 
\left( M_{+}^{\omega _{K}} \right) ^2 E_\pi(\nu,\omega_K )+ 
\left( M_{-}^{\omega _{K}} \right) ^2 E_\nu(\pi,\omega_K )\right] 
\right. \nonumber\\
&& \left. - q^U_{\pi\nu} \sum_{\omega_K} 
M_{+}^{\omega _{K}}  M_{-}^{\omega _{K}}  
\left[ E_\pi(\nu,\omega_K ) +E_\nu(\pi,\omega_K)\right] 
\right\} \nonumber \\
&&+2\kappa ^{pp}_{GT} \left\{ \tilde{q}_{\pi\nu} \sum_{\omega_K}
\left[  M_{+}^{\omega _{K}} M_{++}^{\omega _{K}} E_\pi(\nu,\omega_K )- 
M_{-}^{\omega _{K}} M_{--}^{\omega _{K}} E_\nu(\pi,\omega_K )\right] \right. 
\nonumber\\ 
&&\left. - q_{\pi\nu} \sum_{\omega_K}\left[ 
M_{+}^{\omega _{K}} M_{--}^{\omega _{K}} E_\pi(\nu,\omega_K ) -
M_{-}^{\omega _{K}}  M_{++}^{\omega _{K}} E_\nu(\pi,\omega_K)\right] \right\}\, , 
\label{ampl1qp}
\end{eqnarray}
where $q'$s and $M'$s are given in Eqs. (\ref{m-})-(\ref{qs}).
In this expression, $E_\pi(\nu,\omega_K)$ is in principle given by 
$E_\pi(\nu,\omega_K)=1/(E_\pi-E_\nu-\omega_K)$ and a similar expression for
$E_\nu(\pi,\omega_K)$ changing $\nu$ into $\pi$. Nevertheless, in order to avoid 
accidental singularities, we introduce a width $d=0.5$ MeV in the same way as 
it was done for the first time in Ref. \cite{moll} 

\begin{equation}
E_\pi(\nu,\omega_K )=\frac{E_\pi-E_\nu-\omega_K }{(E_\pi-E_\nu-\omega_K )^2 +
d^2} \, .
\label{epn}
\end{equation}

For simplicity we speak of 1qp, 2qp, and 3qp states, but it should
be clear that we always mean QRPA-correlated states.

Once the intrinsic amplitudes 
$\left< f \left|  \beta ^\pm _{K} \right| i\right>$
are calculated in Eqs. (\ref{ampl3qp})-(\ref{ampl1qp}), 
the Gamow-Teller strength $B_{GT}$ for a transition $I_i \rightarrow I_f$
can be obtained as

\begin{eqnarray}
B^{\pm}_{GT}&=&\sum_{M_i,M_f,\mu} \left| 
\left< I_fM_f \left| \beta ^\pm _\mu
\right| I_i M_i \right> \right|^2 \nonumber \\
&=& \{ \sum_\rho \left[ \left< I_i K_i 1 \rho | I_f K_f \right> \left< 
\phi_{K_f} \left|  \beta ^\pm _\rho \right| \phi_{K_i}\right> 
\right.  \nonumber \\
&&  \left.+(-1)^{I_i-K_i} \left< I_i -K_i 1 \rho | I_f K_f \right> \left< 
\phi_{K_f} \left|  \beta ^\pm _\rho \right| \phi_{\bar{K_i}}\right> 
\right] \} ^2 \nonumber \\
&=& \delta_{K_i,K_f} \left[ \left< I_i K_i 1 0 | I_f K_f \right> \left< 
\phi_{K_f} \left|  \beta ^\pm _0 \right| \phi_{K_i}\right> 
\right. \nonumber \\ 
&&\left. +\delta_{K_i,1/2}(-1)^{I_i-K_i} \left< I_i -K_i 1 1 | I_f K_f \right> 
\left< \phi_{K_f} \left|  \beta ^\pm _{+1} \right| \phi_{\bar{K_i}}\right> 
\right] ^2 \nonumber \\
&& +\delta_{K_f,K_i+1}  \left< I_i K_i 1 1 | I_f K_f \right> ^2 \left< 
\phi_{K_f} \left|  \beta ^\pm _{+1} \right| \phi_{K_i}\right> ^2 \nonumber\\
&& +\delta_{K_f,K_i-1} \left< I_i K_i 1 -1 | I_f K_f \right> ^2 \left< 
\phi_{K_f} \left|  \beta ^\pm _{-1} \right| \phi_{K_i}\right> ^2 \, ,
\label{strodd}
\end{eqnarray}
in units of $g_A^2/4\pi$.
To obtain this expression we have used the initial and final states in the 
laboratory frame expressed in terms of the intrinsic states 
$\left| \phi_K\right>$ using the Bohr-Mottelson factorization\cite{bm}.

Eq. (\ref{strodd}) can be particularized for even-even parent nuclei. In this case
$I_i=K_i=0$, $I_f=1$, and $K_f=0,1$.

\begin{equation}
B^{\pm}_{GT}=\frac{g_A^2}{4\pi}\left\{ \delta_{K_f,0}
\left< \phi_{K_f} \left|  \beta ^\pm _0 \right| \phi_0\right> ^2
+2\delta_{K_f,1}
\left< \phi_{K_f} \left|  \beta ^\pm _1 \right| \phi_0\right> ^2 \right\} \, .
\label{streven}
\end{equation}

\subsection{Excitation energies, $Q_{EC}$ values, and half-lives}

Concerning the excitation energy of the daughter nuclei to which we refer
all the GT strength distributions in this paper, we have to distinguish again
between the case of even-even and odd-A parents.
In the case of even-even systems, this excitation energy is simply given by
\begin{equation}
E_{\mbox{\scriptsize{ex}}\, [(Z,N)\rightarrow (Z-1,N+1)]}=\omega -E_{\pi_0} -
E_{\nu_0} \, , \label{eexeven}
\end{equation}
where $E_{\pi_0}$ and $E_{\nu_0}$ are the lowest quasiparticle energies for 
protons and neutrons, respectively.

In the case of an odd-A nucleus we have to deal with 1qp and 3qp transitions.
Let us consider here our case of an odd-neutron parent decaying by
$\beta^+$ into an odd-proton daughter. 
For 1qp transitions, since the unpaired neutron state is the only neutron 
state involved in the transition, the excitation energy is
\begin{equation}
E_{\mbox{\scriptsize{ex,1qp}}\, (Z,N-1)\rightarrow (Z-1,N)}=E_\pi-E_{\pi_0} \, . 
\label{eex1qp}
\end{equation}
On the other hand, in the 3qp case where the unpaired neutron acts as a spectator, 
the excitation energy with respect to the ground state of the daughter nucleus is 
\begin{equation}
E_{\mbox{\scriptsize{ex,3qp}}\, (Z,N-1)\rightarrow (Z-1,N)}=
\omega +E_{\nu,\mbox{\scriptsize{spect}}}-E_{\pi_0} \, . 
\label{eex3qp}
\end{equation}

This implies that the lowest excitation energy of 3qp type is of the order
of twice the neutron pairing gap. 
In the Kr isotopes under study here, $\Delta \sim 1.5-2$ MeV \cite{sarr2}.
Therefore, all the strength contained in the
low excitation energy region, below typically 3-4 MeV in the odd-A nuclei studied
in this paper must correspond to 1qp transitions. Thus, the low excitation energy 
region (below twice the neutron pairing gap) in a $\beta^+$ decay from an 
odd-neutron nucleus, basically tells us about the proton spectrum.

We should also mention that in odd-A nuclei we have considered the 
transitions to the rotational states as well.
We consider only those states built on the $K_f$ band heads, which are not 
forbidden by the selection rules of the Gamow-Teller operator in the allowed 
approximation. This means that we calculate according to Eq. (\ref{strodd}) 
the GT strength corresponding to transitions from an initial $I_i=K_i$ state 
to final states characterized by
i) $I_f=K_f$ when $K_f=K_i+1$, 
ii) $I_f=K_f$ and $I_f=K_f+1$ when $K_f=K_i$, and
iii) $I_f=K_f$, $I_f=K_F+1$, and $I_f=K_f+2$ when $K_f=K_i-1$.

Accordingly, we have added to the excitation energies $\omega_K$ a standard
Bohr-Mottelson rotational energy \cite{bm}.
The purpose of the inclusion
in our calculation of the transitions to the rotational states is to take into
account in a more reliable way the fragmentation of the GT strength and the
density of excitation energies. Thus, in our calculations we estimate the
moment of inertia by using a mean value between the two extreme macroscopic 
models, rigid rotor (rr) and irrotational flow (if) model, whose predictions are 
usually upper and lower boundary values of the experimental moments of inertia.

$Q_{EC}$ values are also determined differently depending on the even or
odd number of nucleons. The $Q_{EC}$ value is given by 
$Q_{EC}=\left[ M_{\mbox{\scriptsize{parent}}}-M_{\mbox{\scriptsize{daughter}}}
+m_e\right]c^2$, where
$M$'s are the nuclear masses (binding energies of the electrons have been
neglected). 
For the $\beta^+$-decay of an odd-neutron parent we have

\begin{equation}
Q_{EC,(Z,N-1)\rightarrow (Z-1,N)}=m_\pi-m_\nu+m_e+\lambda_{\pi(Z-2,N-2)}-
\lambda_{\nu(Z-2,N-2)}-E_{\pi(Z-2,N-2)}+E_{\nu(Z-2,N-2)}\, ,
\label{qecoddn}
\end{equation}
where $\lambda$ is the Fermi level and $E$ is the lowest 
quasiparticle energy $E=\sqrt{(\epsilon-\lambda)^2+\Delta^2}$.
The $Q$-value for $\beta^+$-decay is simply $Q_{\beta^+}=Q_{EC}-2m_e$.

These expressions can be compared to that corresponding to an even-even
nucleus $(Z,N)$

\begin{equation}
Q_{EC,(Z,N)\rightarrow (Z-1,N+1)}=m_\pi-m_\nu+m_e+\lambda _{\pi,(Z-2,N)}
-\lambda_{\nu,(Z,N)}-E_{\pi,(Z-2,N)}-E_{\nu,(Z,N)}\, .
\label{qeceven}
\end{equation}

The half-lives for the $\beta^+/EC$-decay are obtained using the standard 
definition
(see for instance Ref.\cite{sarr2}), involving sums over all possible final 
states within the $Q-$window reached in the decay and including standard 
quenching factors \cite{oster}.

The Fermi integrals required for computation of half-lives have been  obtained 
numerically for each value of the charge of the daughter nucleus $Z$ 
and the maximum energy available $W_0$ in $\beta$-decay, as explained in 
Ref.\cite{gove}.

\section{Results}

In this Section we present and discuss the results obtained for the GT strength
distributions, half-lives, and summed strengths in the proton rich Kr isotopes. 
The results correspond to QRPA
calculations with the Skyrme force SG2 and they have been performed for the nuclear
shapes that minimize the HF energy. 
Before discussing the figures we note that the GT strength distributions 
are plotted versus
the excitation energy of the daughter nucleus.
The distributions of the GT strength in Figs. 3, 5, and 6 have been folded 
with $\Gamma =1$ MeV width Gaussians to facilitate the comparison among the 
various calculations, so that the original discrete spectrum is transformed 
into a continuous profile.
The distributions in those figures are given in units of $g_A^2/4\pi$ and 
one should keep in mind that a quenching of the strength 
is expected on the basis of the observed quenching in charge exchange reactions 
and spin $M1$ transitions in stable nuclei, where 
$g_{s,{\mbox{\scriptsize{eff}}}}$ is also known to be approximately 
$0.7\ g_{s,{\mbox{\scriptsize{free}}}}$. Therefore, a reduction factor of 
about two is expected in these strength distributions in order to compare 
with experiment.
This factor is of course taken into account when comparison to experimental 
half-lives, summed strengths, and GT strengths is made.

\subsection{Effect of the residual interaction}

We can see in Fig. 3 the effect of the residual interaction treated in QRPA
on the uncorrelated calculation (dotted lines denoted by MF). We do that on 
the example of $^{73,74}$Kr for the oblate and prolate shapes that minimize 
the energy obtained with the Skyrme force SG2. The coupling strengths of the 
$ph$ and $pp$ residual interactions are $\chi ^{ph}_{GT}=0.37$ MeV  
and $\kappa ^{pp}_{GT}=0.07$ MeV, respectively.
Both $ph$ and $pp$ residual interactions reduce the GT strength. The residual
forces produce also a displacement of the GT strength, which is to higher
energies in the case of the repulsive $ph$ force and to lower energies in
the case of the attractive $pp$ force.

These effects are common to even-even and odd-A isotopes.
Nevertheless, we observe that the effect of the residual interaction in 
odd-A nuclei is very small  in the low energy tail of the GT strength
distribution. This is especially true for the $pp$ force. The 
reason for that can be understood from the fact that the lowest-lying
transitions are 
affected by the residual force only through the weak correlations with 
phonons treated in first order perturbation.

This feature has also important consequences when one considers the 
half-lives because they depend only on the distribution of the strength below 
the $Q_{EC}-$energy window. Therefore, the $pp$ interaction will affect the 
half-lives differently if we deal with even-even or odd-A nuclei. 
In Fig. 4 we have plotted the half-lives of the $^{71,72,73,74}$Kr isotopes 
as a function of the strength  $\kappa ^{pp}_{GT}$. As we can see from this
figure the half-lives decrease with increasing values of  $\kappa ^{pp}_{GT}$,
but in a different way depending on the isotopes. We can see that the
half-lives of odd-A nuclei have a very smooth, almost flat, behavior with 
the $pp$ force, while the half-lives of even-even nuclei present a stronger 
dependence on the $pp$ force. Decreasing of half-lives with 
$\kappa ^{pp}_{GT}$ can be understood from the fact that the $pp$ force is 
attractive and therefore tends to concentrate the strength to lower energies. 
The reason for the smoother decrease of half-lives in the odd-A 
isotopes is related to the smaller effect of the $pp$ force in the low-energy
tail of the GT strength distribution in odd-A nuclei, previously discussed.

The optimum value of $\kappa ^{pp}_{GT}$  to reproduce the half-life depends, 
among other factors, on the nucleus, shape, and Skyrme interaction and a case 
by case fitting procedure could be carried out. Since the half-lives are
practically insensitive to this force in odd-A nuclei, this fit could be
restricted to even-even nuclei. In Ref. \cite{sarr4} we considered this 
dependence in various even-even isotopes in the mass region $A\sim 70$
(Ge, Se, Kr, and Sr), and arrived to the conclusion that a value of 
$\kappa ^{pp}_{GT}=0.07$ MeV improves the agreement with experiment in 
most cases. Thus, we also use in this paper the same coupling strength 
for the $pp$ force.

\subsection{Comparison of even-even and odd-A GT strengths}

To further illustrate the relative importance of the different types of 
contributions to the GT strength in odd-A nuclei, we show in Fig. 5  the GT
strengths (solid) decomposed into their 1qp (dotted) and 3qp (dashed) 
contributions. 
The picture emerging from the analysis of this figure is that
the GT strength distributions in odd-A nuclei can be divided into two
different regions. One is the energy region below twice the pairing gap, 
where the individual excitations are determined by the quasiparticle proton 
(neutron) energies in the case of an odd-neutron (odd-proton) parent nucleus. 
This region is of relevance for
$\beta^+$-decay since it appears within the $Q$-window. The other region
at higher energies is dominated by 3qp excitations, where the odd-nucleon
acts as a spectator. The strength contained here is much larger than in
the low energy region because many more configurations are possible but only
in the very proton rich isotopes this is accessible by $\beta^+$-decay.

Fig. 6 contains a summary of the results on GT strengths obtained in this work. 
In this figure 
we can see our HF+BCS+QRPA Gamow-Teller strength distributions predicted by
the Skyrme force SG2 for the whole Kr isotopic chain including odd-A and 
even-even nuclei. 

The trend observed in the GT strength distributions of both odd-A and 
even-even isotopes is similar. We can see that the most unstable isotopes
have the largest strengths which are also placed at higher energies.
As we move into the stable isotopes by increasing the number of neutrons, 
the GT resonance appears at lower excitation energies and contains less and
less strength. Also the $Q_{EC}$ window, represented by the vertical line, 
becomes smaller and smaller.

We can observe also the similarity between the even-even and odd-A partners.
If we compare the strengths for an even-even $(N,Z)$ nucleus with that of the
corresponding odd-A $(N-1,Z)$ we can see that they are about the same once the
low energy region of about twice the neutron pairing gap (between 3 and 4 MeV 
depending on the case) is suppressed. 

We can also observe that the different profiles of the GT strength 
distributions corresponding to the various shapes of a given isotope
can be used in certain cases as a signature of the nuclear shape.

\subsection{Low-energy GT strength and comparison to experiment}

Now we concentrate on the energy region below $Q_{EC}$ and 
discuss the possibilities to discriminate between different shapes by 
$\beta^+$-decay experiments. We also compare with the available experimental 
data. We perform this detailed analysis by plotting not the folded strengths
as it was done in the previous figures, but the individual excitations as
they come from the calculation. The GT strengths have been quenched with the
same factor used to calculate the half-lives. 
In Figs. 7-9 we can see these results for the even-even isotopes  
$^{72,74,76}$Kr, respectively. 

Fig. 7 for $^{72}$Kr includes the results from our QRPA calculations with
the force SG2 with oblate and prolate shapes. Below an excitation energy of
about 2 MeV we can see that the distribution of the GT strength predicted
by the oblate or prolate shapes is qualitatively very similar although  
the prolate shape gives somewhat larger strength. We obtain peaks at 0.5 MeV
and 1.5 MeV  in both cases and they are in agreement with preliminary data
on this nucleus \cite{private}. Therefore, it will be hard to distinguish
between the two shapes.  
On the other hand, if we look into the energy range from 2 MeV up to $Q_{EC}$, 
we can see a strong double peak that appears in
the oblate case between 2 and 2.5 MeV. The strength in this region is about 
three times larger than the strength of the first peak at 0.5 MeV and it is
almost absent in the prolate case. The appearance or absence of this peak
at 2 MeV could be the signature of an oblate or prolate shape, respectively.
It is also worth mentioning the huge peak appearing in the prolate case very
close to the $Q_{EC}$ value. If it could be seen experimentally, it would be
a clear signature in favor of a prolate shape. 
We can see in Table 1 the total measured GT strength below 1.836 MeV 
\cite{exp72} compared to our results with the two shapes, where oblate shape
seems to be favored.

Fig. 8 shows the results in $^{74}$Kr. Experimental data are from Ref.
\cite{exp74}.  We can distinguish two regions in this isotope.
At energies below 2 MeV we find that the oblate shape predicts much more 
strength than the prolate shape and the opposite happens beyond 2 MeV.
Therefore, measuring the GT strength distribution in  $^{74}$Kr up to 
$Q_{EC}$ would help to discriminate between the two shapes. 
If the strength is concentrated
below 2 MeV it will correspond to an oblate shape, while if it is concentrated 
between 2 and 3 MeV it will correspond to a prolate shape. A comparison
of the strength contained below 1 MeV is made in Table 1. The experimental 
summed strength \cite{exp74} lies between the predictions of the two shapes 
but it is closer to the strength produced by the oblate shape.

In Fig. 9 we show the results for  $^{76}$Kr corresponding to the spherical
and prolate shapes that minimize the energy in this nucleus. Experimental data
are from Ref. \cite{exp76}. 
In this case, an isolated single peak in 
the measured strength at very low excitation energy would be the signature
of a spherical parent, while a peak close to the $Q_{EC}$ limit would be
the signature of a prolate parent. Comparison with the available data seems
to favor the spherical shape. This is also true if we compare the total
GT strengths contained below 1 MeV, as can be seen in Table 1, where the
strength generated by the spherical shape is much closer to experiment.

Next figures 10-13 correspond to the odd-A nuclei  $^{73,75}$Kr. In Fig. 10
we can see our results for the oblate and prolate shapes of  $^{73}$Kr
compared to the experimental data from Ref. \cite{exp73}. The first thing
to mention is that the experimental data have been taken differently in the
two energy ranges below and above 3.5 MeV. Below 3.5 MeV the data have been
extracted from direct detection of the gamma rays. Beyond this energy they
have been extracted from proton delayed detection and have big
errors. A new effort to measure experimentally all the energy range up to
$Q_{EC}$ with large efficiency gamma ray detectors
is under way at ISOLDE\cite{isolde}. Until these data become available and 
confirm the present measurements, the data measured above 3.5 MeV should 
be considered as partial because part of the strength escapes observation
due to the high level density \cite{private}.
The summed GT strengths in the two energy regions can be seen in Table 1.
The experimental data appear between the predictions of the oblate and prolate
shapes, the oblate one being larger.

Comparison to the data of the GT strength distributions below 3.5 MeV shows
that both oblate and prolate shapes produce similar strength in the very low
excitation energy range below 0.5 MeV, which is compatible with experiment.
At higher energies the prolate shape does not generate strength up to 4 MeV
while the oblate shape generates two bunches at 1 MeV and 3.5 MeV that
compare better with experiment. Above 4 MeV, the strength increases and 
the prolate shapes produces two bumps at 4.5 and 6 MeV, while the oblate 
shape produces a wide peak from 4.5 to 6.5 MeV with a huge strength at an
excitation energy of 5.5 MeV. 
It is also worth mentioning that the structure of the GT transitions in the
theoretical calculations can be analyzed taking into account the rotational
nature of the final states reached by the allowed GT transition. As we
have already mentioned, since in this case we are considering the transitions
$K_i^\pi=3/2^- \rightarrow  K_f^-$ with $K_f=1/2,3/2,5/2$, when we reach a
final state with $K_f^\pi=1/2^-$, we consider the rotational states
with $I_f=1/2^-,3/2^-,5/2^-$. The strengths of these states are given by the
geometrical Clebsch-Gordan factors in Eq. (\ref{strodd}) and their excitation
energies by the rotational energies. Similar arguments apply to the transitions to
$K_f^\pi=3/2^-$, where we consider the rotational states with $I_f=3/2^-,5/2^-$.
For $K_f^\pi=5/2^-$, the only case to account for is $I_f=5/2^-$.
On this basis we can now see that the two peaks below 0.5 MeV in the prolate 
case are the members of a rotational band $I_f=K_f=3/2^-$ and  
$I_f=5/2^-,K_f=3/2^-$. Similarly we can identify the low-lying structure in the
oblate case as the members of several rotational bands. This can be seen more
easily in Fig. 11, where we have plotted the excitation energies in the
daughter nucleus $^{73}$Br  of the states reached by $\beta^+$-decay. In this
figure we have also added by dotted lines the first forbidden transitions 
that correspond to
$\Delta J=1$ with parity change, that is, states $1/2^+, 3/2^+, 5/2^+$.
Although we have not calculated their GT strength, their excitation energies
have been plotted in the figure. We can see that the agreement with experiment
is very reasonable in the oblate case. We get concentrations
of states at the same energies and the total number of states is similar.
In this figure we can follow more easily the structure of the bands. For example,
in the prolate case we can see that the lowest $3/2^-$ has a $5/2^-$ associated,
the next $1/2^-$ has two states $3/2^-$ and $5/2^-$ associated and so on.

A similar analysis has been done in Figs. 12-13 for $^{75}$Kr. In Fig. 12 we
can see the GT strengths in the prolate and spherical cases compared to
the experimental data from Ref. \cite{exp75} taken up to 2.5 MeV. 
The summed strengths can be also compared in Table 1. The spherical shape
does not produce any strength in the $Q_{EC}$ window except a small bunch
of states around 3 MeV. On the other hand the prolate shape generates strength
at the right position although a little bit smaller than experiment. This 
shape also predicts some strength close to $Q_{EC}$, which is not present
in the spherical case. Fig. 13 shows the experimental low-lying  energy 
spectrum compared to our calculation for the prolate shape. In our results
we can identify the origin of these excitations. We have a low-lying 
$3/2^+,5/2^+,7/2^+$ rotational triplet and a $5/2^+,7/2^+$ doublet. They
correspond well with the experimental energies. We should remember that
the spacing in the rotational energies is determined by the moment of inertia,
and we are using a rough estimate. Using a moment
of inertia a little bit larger, the energy levels become compressed and 
agree better with the experimental spacing. We have also added by dotted
lines the energies of first forbidden transitions.

Experimental information on the decay of $^{71}$Kr is available for the
low-lying states \cite{oino}. In that reference, final states for GT decay
were reported to the ground state of $^{71}$Br and to excited states at
9 and 207 keV. However, the spin-parity assignment of these states, as well 
as the assignment to the  ground state of the parent nucleus $^{71}$Kr are 
still controversial \cite{urkedal}. 
We can see in Fig. 14 the results of our mean field calculations with the
force SG2 for the oblate and prolate shapes of $^{72}$Kr. The vertical axis
is the neutron occupation probability and the horizontal one is the neutron 
single-particle energy. The deformed single-particle states are labelled by
$K^{\pi}$. We can see that the picture is compatible with several spin and parity
assignments for the odd-neutron isotope $^{71}$Kr. 
If we consider the neutron level closest to Fermi level in Fig. 14, the 
spin-parity assignment would be $9/2^+$ for the oblate shape and $3/2^-$ for 
the prolate shape. On the other hand, the assumption taken in Ref. \cite{oino}
was $5/2^-$, which is also close to Fermi level. We therefore consider these
two possibilities for each shape.

We show in Fig. 15 the results for the GT strengths from our calculations 
assuming the above mentioned possibilities for the spin and parity
of the parent nucleus, $5/2^-$ as it was taken in Ref. \cite{oino}, as well as
$9/2^+$ in the oblate case and $3/2^-$ in the prolate case. We can also see 
the spin and parity of each GT excitation.

\subsection{$Q_{EC}$ values and half-lives}

Experimental $Q_{EC}$ and $T_{1/2}$ values are plotted as circles in Fig. 16 for 
the Krypton isotopes considered in this work. The data have
been taken from Ref. \cite{audi}, except for the isotopes $^{70,71}$Kr, where
we have used the more recent data from Ref. \cite{oino}.
These values are compared to our theoretical results represented by the 
vertical lines. The extreme values of these vertical lines correspond to 
the results obtained from the various shapes using the forces SG2 and Sk3.
This has been done in order to have a better idea of the theoretical spread 
of the results. The agreement is in general good for both $Q_{EC}$ and 
half-lives. This is notorious because there is a very large range of variation
(seven orders of magnitude) for half-lives.
$Q_{EC}$ values are well reproduced with the exception of $^{70}$Kr, where
we obtain a value below experiment. 

We can also see that the calculations fail to account for the half-lives of the
most unstable 
nuclei $^{70}$Kr and $^{71}$Kr. Nevertheless, it should be mention that the
calculations correspond to GT transitions neglecting possible contributions
from Fermi transitions. But Fermi transitions might play a significant role
to obtain the total decay rates in nuclei where $N\sim Z$. Indeed, we have 
calculated the Fermi strength distributions and evaluated the corresponding
half-lives. The calculation of Fermi transitions follows closely that of 
Gamow-Teller transitions. One should simply use an isospin-isospin residual
interaction with a coupling strength derived from the Skyrme force as shown
in Ref. \cite{sarr1} and replace the spin matrix elements between proton 
and neutron states given in Eq. (\ref{sigma}) by the overlaps $< \nu|\pi >$. 
Of course, only $K=0$ components will survive. 

We find that the half-lives corresponding to Fermi transitions are negligible
as compared to the GT half-lives except in the isotopes $^{70,71,72,73}$Kr,
where they are comparable and therefore one should not neglect their 
contribution. We can see in Fig. 16, the new results for the total half-lives
once the contribution from Fermi transitions have been included. They are
given by the vertical lines located at the right side in the half-lives
of the $^{70,71,72,73}$Kr isotopes. The agreement with experiment improves in 
the cases of $^{70,71}$Kr but we are still above the experiment. Nevertheless,
it should be mentioned that if we use the experimental $Q_{EC}$ value in $^{70}$Kr,
instead of the $Q_{EC}$ result from our calculations, we get agreement with 
the experimental half-life. Thus, the discrepancy found in the half-life
of  $^{70}$Kr is entirely due to the discrepancy in the $Q_{EC}$ value.

We can observe in Fig. 16 a nice trend that can be followed by the auxiliary
dotted lines joining the experimental data of even-even isotopes on one hand
and the odd-A on the other. The growing $Q_{EC}$ trends in the two lines of 
Fig. 16 are dictated by the increase of ($\lambda_\pi -\lambda_\nu$) as one
approaches the proton drip line.
This behavior can be qualitatively understood in the case of $Q_{EC}$ from
their expressions given by Eqs. (\ref{qecoddn})-(\ref{qeceven}).
As we have mentioned earlier, the only exception to this trend appears in
the experimental $Q_{EC}$ value of $^{70}$Kr, which is clearly above the 
theoretical value.
Since in this case, the daughter nucleus $^{70}$Br is odd-odd $N=Z$, 
one may argue that the
observed jump from the regular trend in $Q_{EC}$ and $T_{1/2}$ is due to an 
extra neutron-proton binding, which is not taken into account in the present 
calculation.

\section{Conclusions and final remarks}

We have applied a selfconsistent deformed HF+BCS+QRPA formalism with
density-dependent effective Skyrme interactions to the description of 
the $\beta$-decay properties of proton-rich odd and even Kr isotopes. This 
approximation
has the appealing feature of treating the excitations and the ground state 
in a selfconsistent framework with basically no free parameters.
Our spin-isospin residual interaction contains a particle-hole part, which 
is derived selfconsistently from the Skyrme force, and a particle-particle 
part, which is a separable force representing a neutron-proton pairing force.

We have analyzed the similarities and differences in the treatment and
in the results of even-even and odd-A nuclei. The low-lying GT response 
of odd-A nuclei is generated by transitions involving the state of the odd nucleon 
(1qp transitions). Thus, the excitation spectrum up to twice the pairing gap
parameter of neutrons (protons) in the $\beta^+$-decay of and odd neutron 
(proton) parent, gives information on the proton (neutron) states. 
When this low energy strength in the odd-A nuclei is removed, the 
resulting GT strength distribution is very similar to that of the even-even 
neighbor displaced to higher excitation energy $(E_{ex}\sim 2\Delta)$.

In odd-A deformed nuclei, for each allowed intrinsic GT transition, one has a 
set of transitions to rotational states with decreasing strength at higher 
energies. This is a characteristic of odd-A deformed nuclei,  which is not 
present in the spherical limit.  The GT strength corresponding to transitions 
to states in a given 
band is reduced by geometrical factors involving angular momenta. The energy 
separation depends on the angular momentum of the transition with a global 
scale determined by the moment of inertia.

We have found a reasonable agreement with available experimental data.
We have also discussed what can be learned from future comparison of our 
results with experimental data from ISOLDE that are expected very soon.
This comparison  would be interesting for several reasons:
i) From our study of the dependence on the shape of the GT strength 
distributions we conclude that information on the shape of the parent 
nucleus can be gained when data in the whole $Q_{EC}$ window become available.
We have identified particular narrow regions in the excitation spectra of
various nuclei where data could be more conclusive on the nuclear shape; 
ii) Comparison to data would be desirable before considering further 
theoretical refinements, such as the effects of the continuum, the 
neutron-proton pairing at the mean field level or extensions of the RPA;
iii) An important point will be to check whether our calculations produce
the same level of agreement in all the isotopes or some special characteristics
can be found in $N=Z$ nuclei.
For the moment, an interesting feature found is the deviation between theory and
experiment in the half-life of $^{70}$Kr. From the present calculation, this
deviation can be attributed to the difference between experimental and theoretical
$Q_{EC}$ values, which in turn can be a signature of an extra binding in the
$N=Z$ odd-odd daughter nucleus.

\acknowledgments 

We are thankful to M.J.G. Borge, Ch. Mieh\'e, W. Gelletly and E. Roeckl for 
stimulating comments and discussions.
This work was supported by DGESIC (Spain) under contract number PB98-0676.

\newpage

\begin{table}[tbp]

{\bf Table 1.} Comparison of the GT strengths contained below some given 
excitation energy between experimental measurements (\cite{exp72} for $^{72}$Kr, 
\cite{exp73} for $^{73}$Kr, \cite{exp74} for $^{74}$Kr, \cite{exp75} for 
$^{75}$Kr, and \cite{exp76} for $^{76}$Kr), and theoretical calculations.
\vskip 0.5cm
\begin{tabular}{lccc}
& exp & oblate & prolate \\ 
$^{72}$Kr ($E_{ex} \leq $ 1.836 MeV) & 0.5 $\pm$ 0.1 & 0.5 & 0.8 \\
$^{73}$Kr ($E_{ex} \leq $ 3.5 MeV) & 0.10 $\pm$ 0.02 & 0.19 & 0.06 \\
$^{73}$Kr (4.0 MeV $\leq E_{ex} \leq $ 6.5 MeV) & 0.83 $\pm$ 0.60 & 0.98 & 0.42 \\
$^{74}$Kr ($E_{ex} \leq $ 1 MeV) & 0.20 & 0.23 & 0.12 \\
$^{75}$Kr ($E_{ex} \leq $ 2.2 MeV) & 0.08 & 0.00 (sph) & 0.05 \\
$^{76}$Kr ($E_{ex} \leq $ 1 MeV) & 0.16 & 0.18 (sph) & 0.05 
\end{tabular}
\end{table}

\newpage

\begin{figure}[t]
\psfig{file=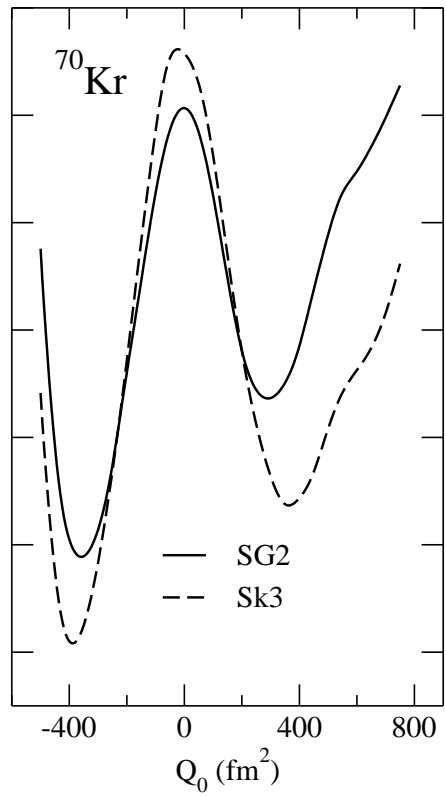,width=0.8\textwidth}
\caption{ Total energy of $^{70}$Kr as a function of the mass quadrupole 
moment $Q_0$ obtained from a constraint HF+BCS calculation with the Skyrme 
forces SG2 (solid) and Sk3 (dashed). The distance between ticks in the vertical
axis corresponds to 1 MeV but the origin is different for the two forces.}
\end{figure}
\vfill\eject

\begin{figure}[t]
\psfig{file=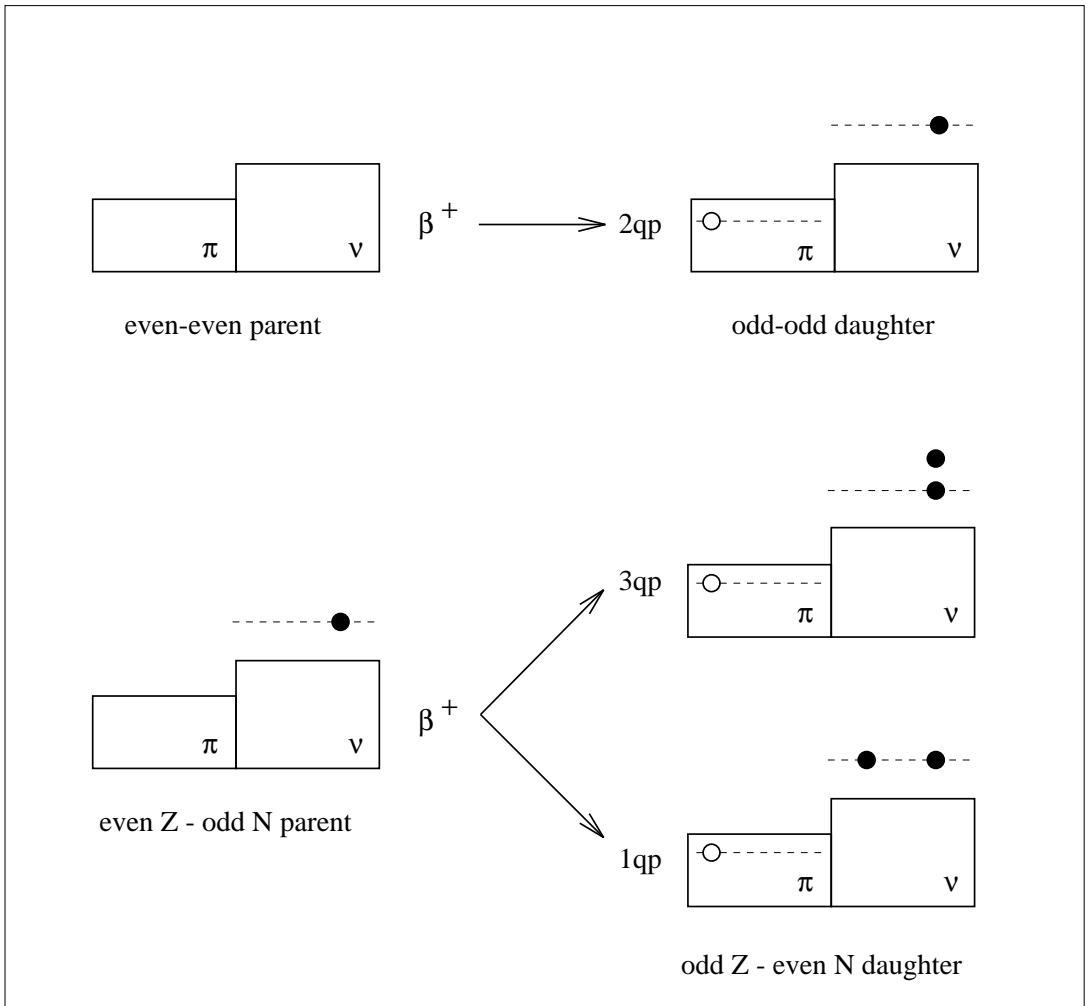,width=0.8\textwidth}
\caption{ Schematic picture to illustrate the different types of
$\beta^+$ decay in the extreme single-particle model. In the case of an 
even-even parent we have 2qp transitions. In the case of an odd neutron parent
there are two types of transitions.
In the 3qp case the unpaired neutron in the parent nucleus acts as
an spectator. The 1qp type of transitions are those involving the unpaired
neutron.}
\end{figure}
\vfill\eject
\begin{figure}[t]
\psfig{file=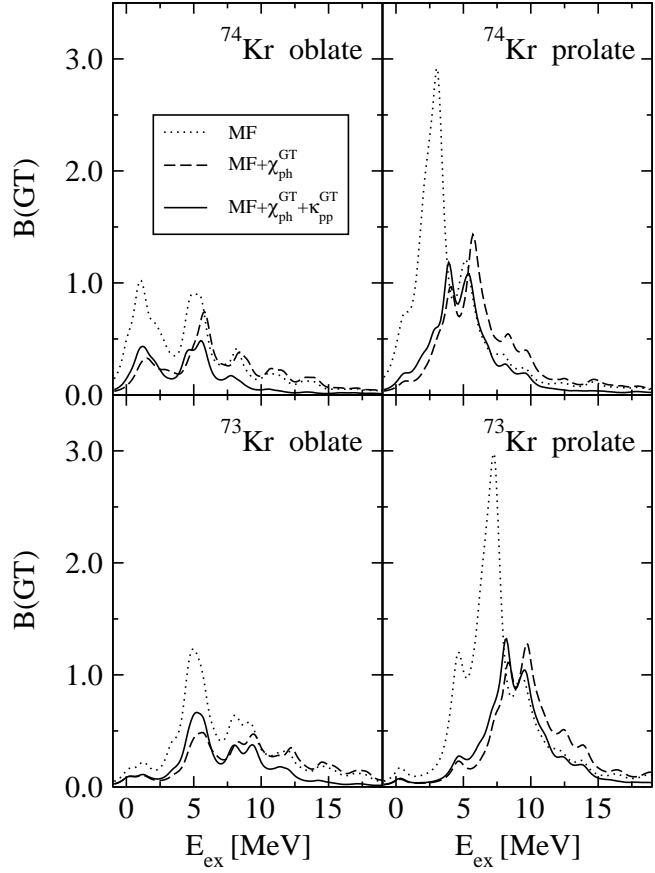,width=0.8\textwidth}
\caption{ Gamow-Teller strength distributions $[g_A^2/4\pi]$ in 
$^{73,74}$Kr isotopes plotted versus the excitation energy of the 
corresponding daughter nucleus $^{73,74}$Br, respectively. The calculations
are performed in HF+BCS approximation (dotted lines denoted by MF)
and in QRPA including only the $ph$ residual interaction (dashed) and including
both $ph$ and $pp$ residual interactions (solid).}
\end{figure}
\vfill\eject
\begin{figure}[t]
\psfig{file=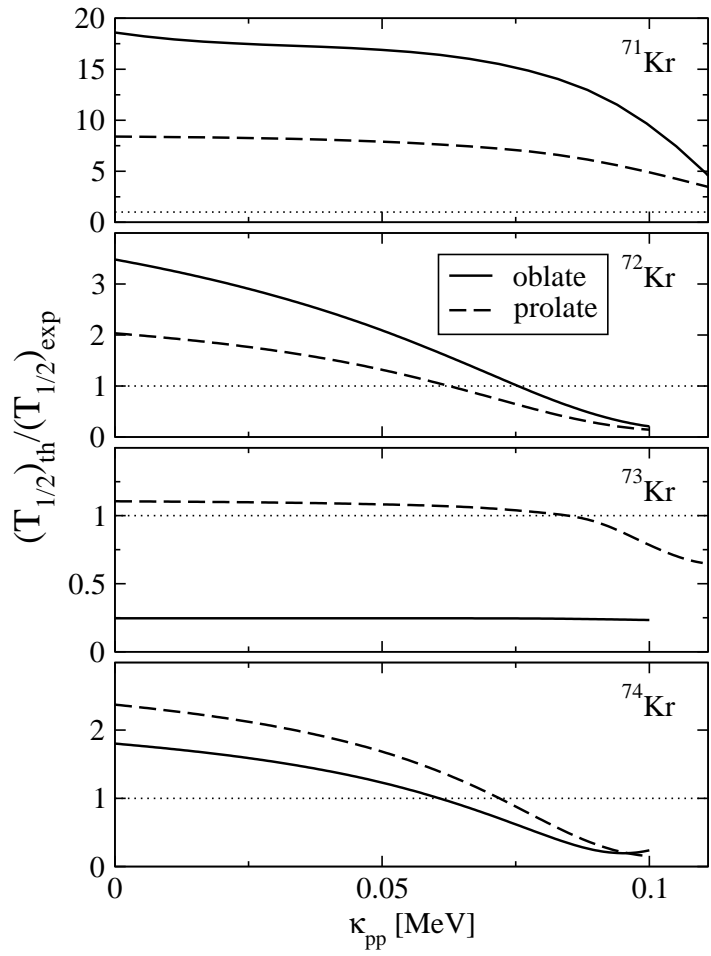,width=0.8\textwidth}
\caption{  Ratios of calculated to experimental half-lives in Kr isotopes
as a function of the coupling strength of the $pp$ force.}
\end{figure}
\vfill\eject
\begin{figure}[t]
\psfig{file=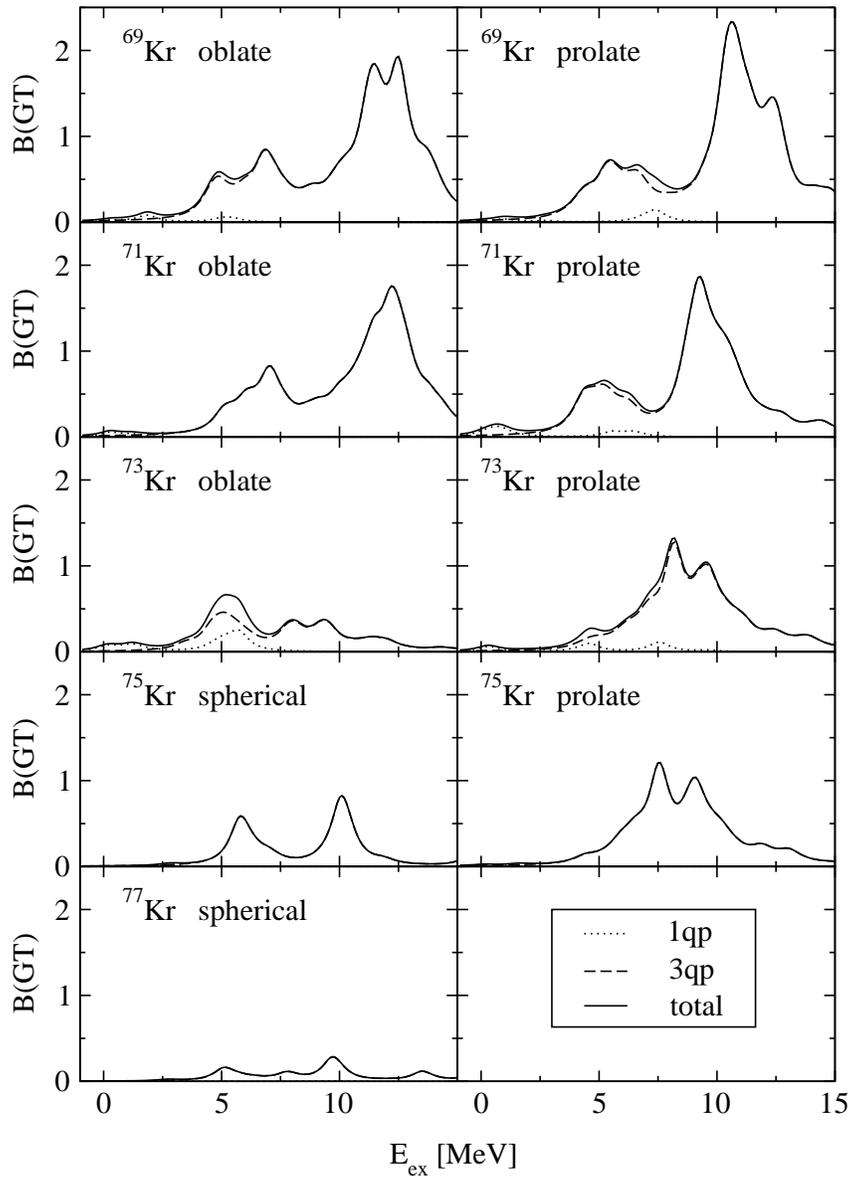,width=0.8\textwidth}
\caption{ Decomposition of the total GT strength distribution into their
1qp and 3qp contributions (see text).}
\end{figure}
\vfill\eject
\begin{figure}[t]
\psfig{file=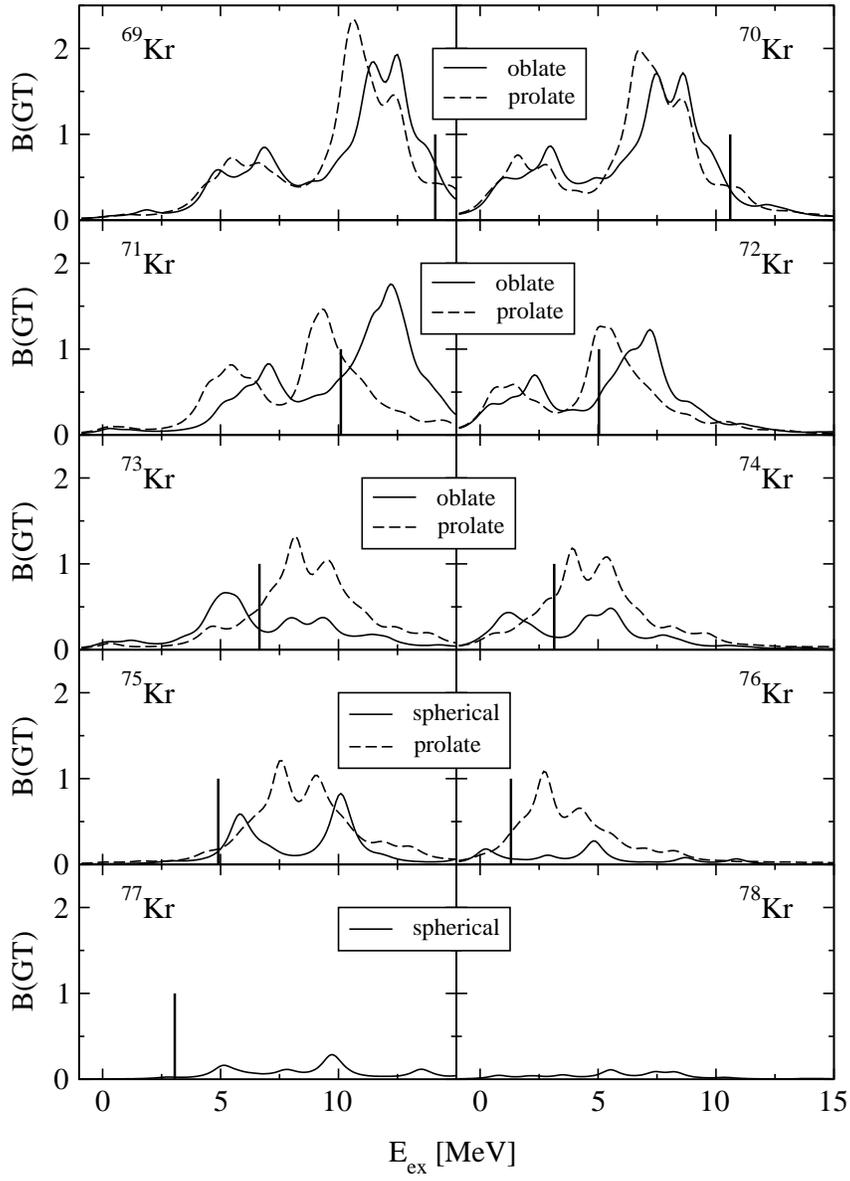,width=0.8\textwidth}
\caption{ Gamow-Teller strength distributions $[g_A^2/4\pi]$ as a function
of the excitation energy of the daughter nucleus [MeV]. The results correspond
to the Skyrme force SG2 in QRPA for the various shapes of the even-even and
odd-A Krypton isotopes.}
\end{figure}
\vfill\eject
\begin{figure}[t]
\psfig{file=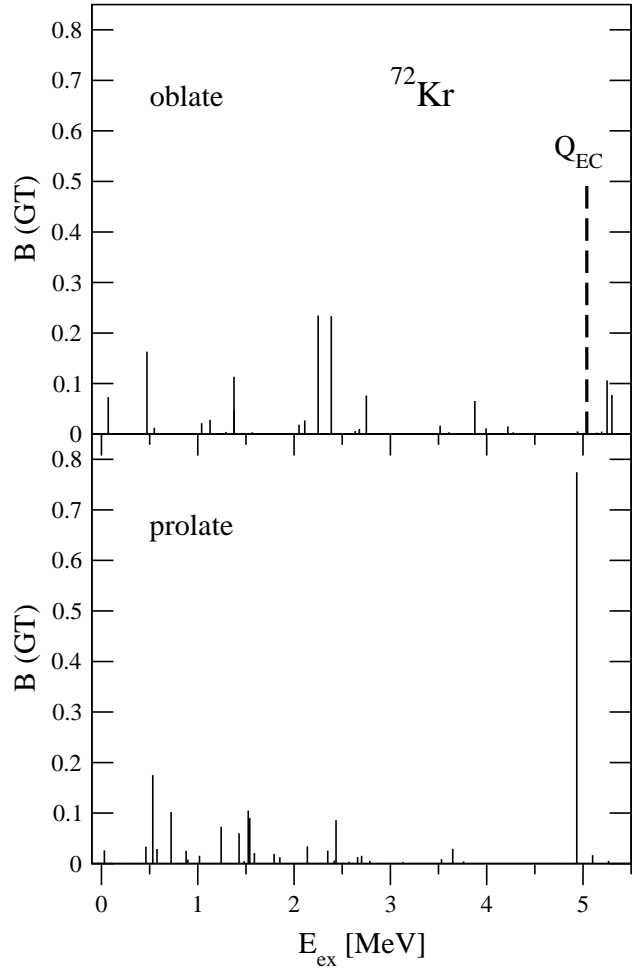,width=0.8\textwidth}
\caption{ Gamow-Teller strength transitions in the $\beta^+$-decay of
$^{72}$Kr as a function of the excitation energy of the daughter nucleus
$^{72}$Br.}
\end{figure}
\vfill\eject
\begin{figure}[t]
\psfig{file=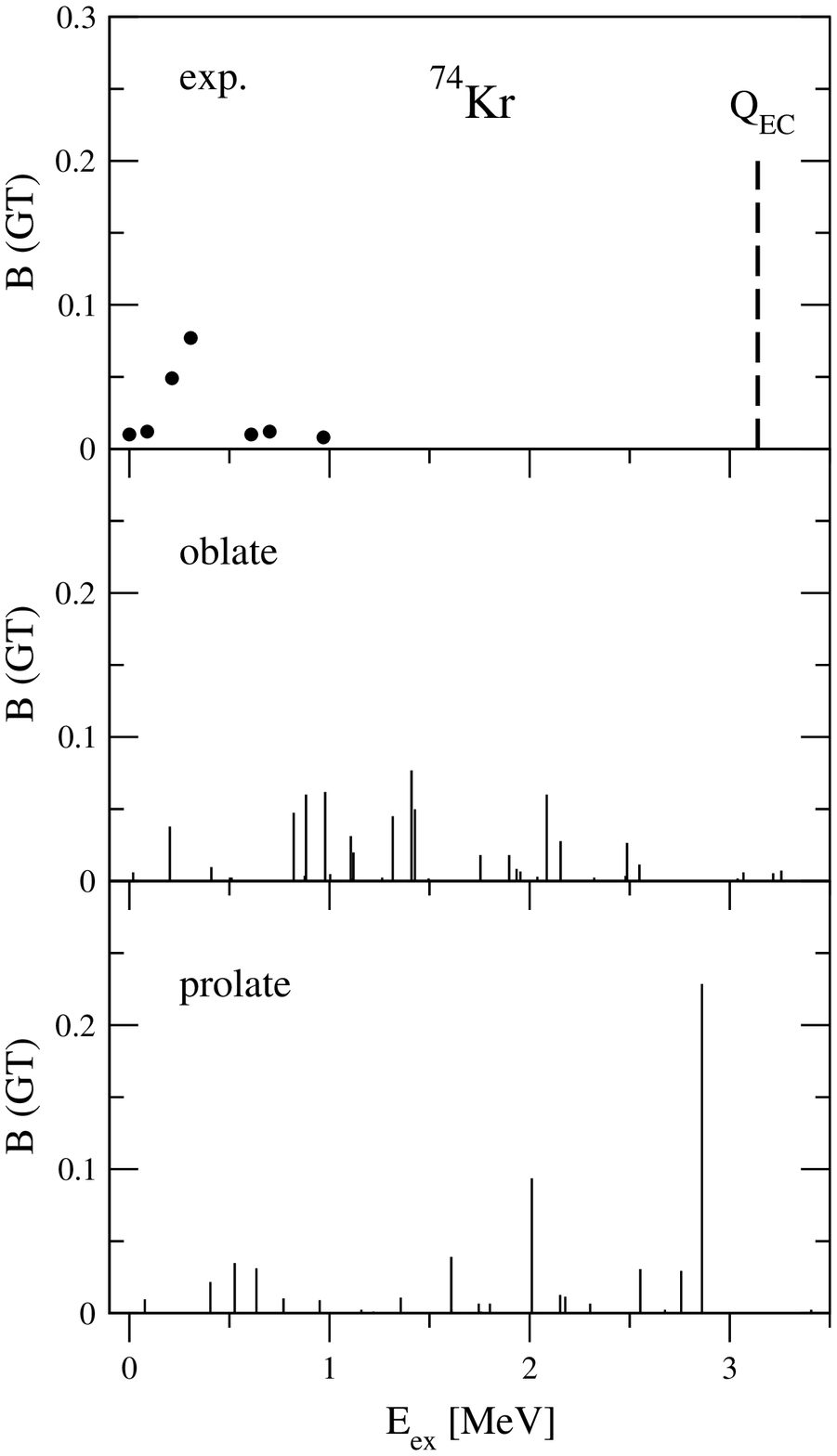,width=0.8\textwidth}
\caption{ Same as in Fig. 7 for the decay of  $^{74}$Kr. 
Experimental data are from [33].}
\end{figure}
\vfill\eject
\begin{figure}[t]
\psfig{file=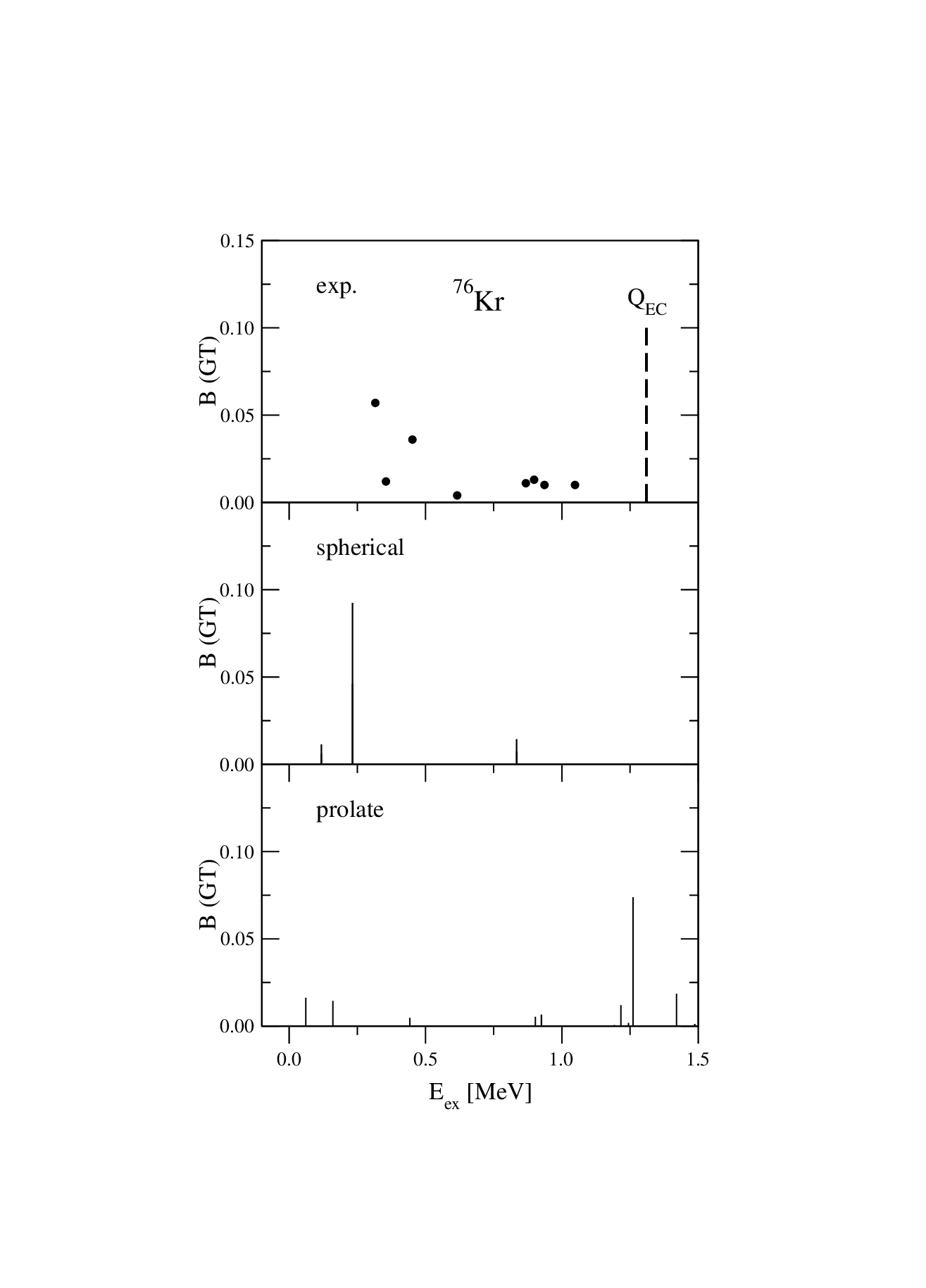,width=0.8\textwidth}
\caption{  Same as in Fig. 7 for the decay of  $^{76}$Kr.  
Experimental data are from [34].}
\end{figure}
\vfill\eject
\begin{figure}[t]
\psfig{file=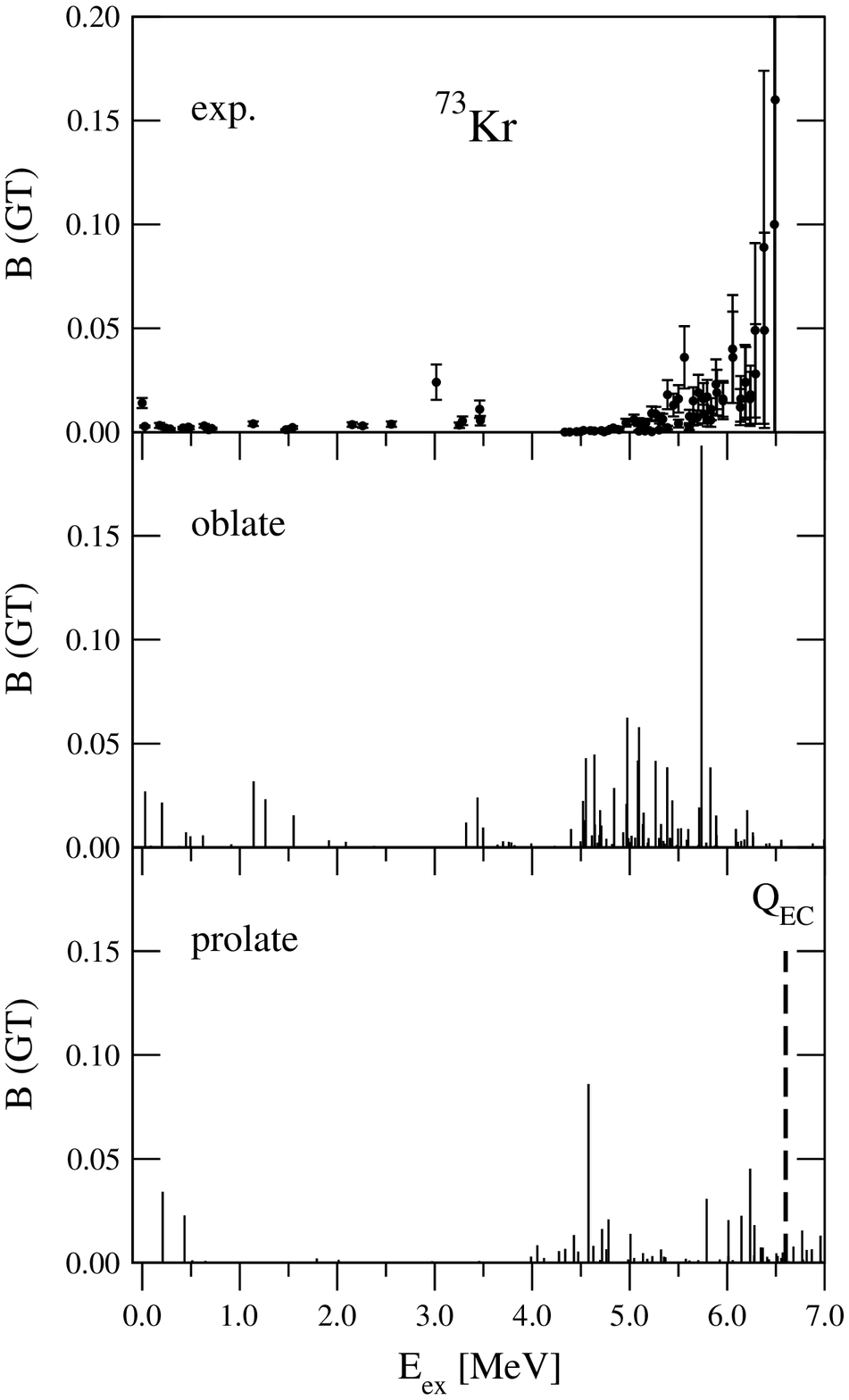,width=0.8\textwidth}
\caption{  Same as in Fig. 7 for the decay of  $^{73}$Kr.
Experimental data are from [23]. }
\end{figure}
\vfill\eject
\begin{figure}[t]
\psfig{file=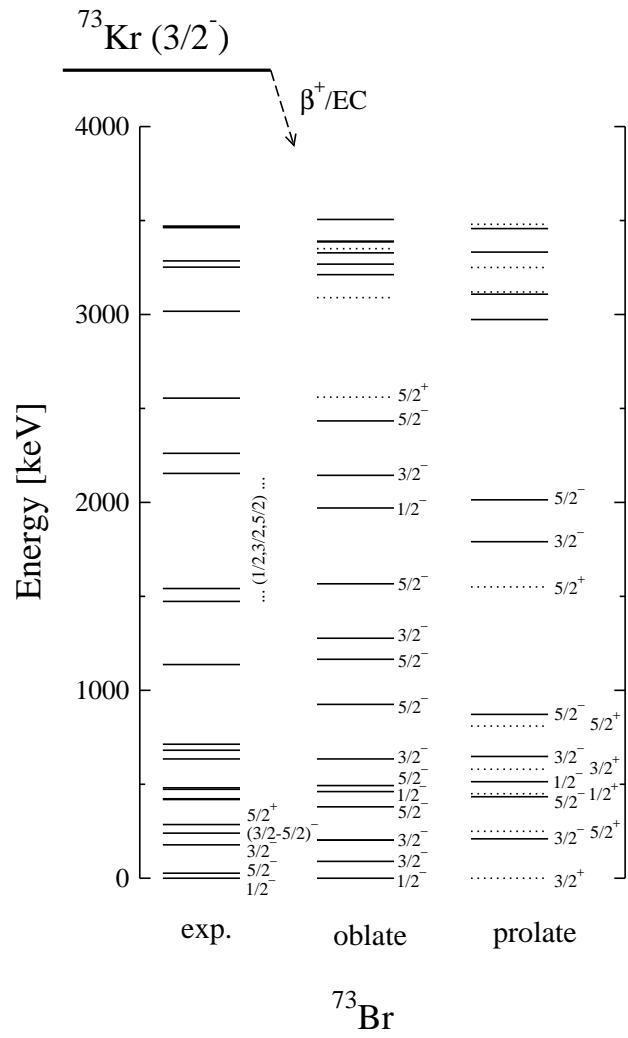,width=0.8\textwidth}
\caption{ Experimental and calculated decay schemes for $^{73}$Kr. }
\end{figure}
\vfill\eject
\begin{figure}[t]
\psfig{file=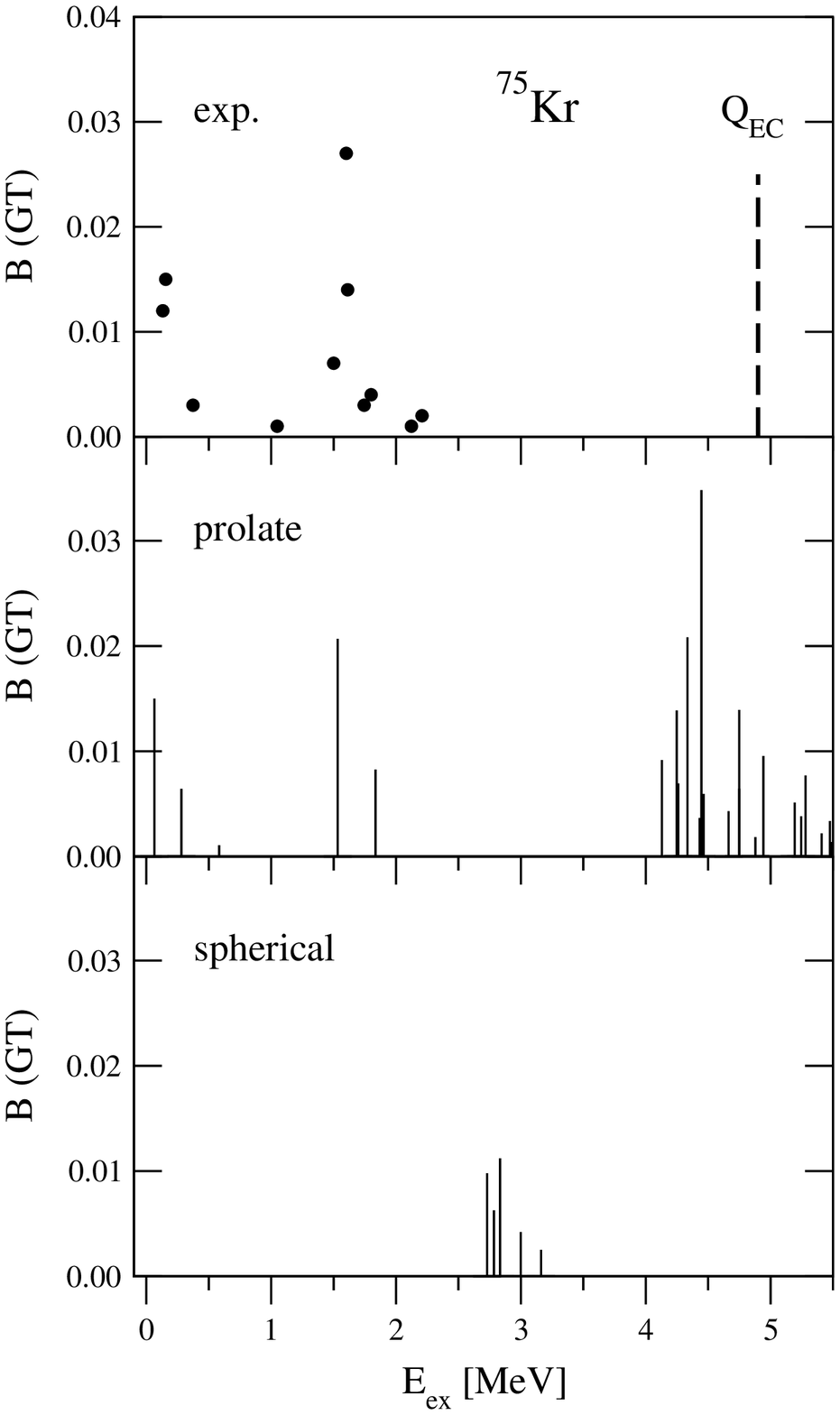,width=0.8\textwidth}
\caption{ Same as in Fig. 7 for the decay of  $^{75}$Kr. 
Experimental data are from [36].}
\end{figure}
\vfill\eject
\begin{figure}[t]
\psfig{file=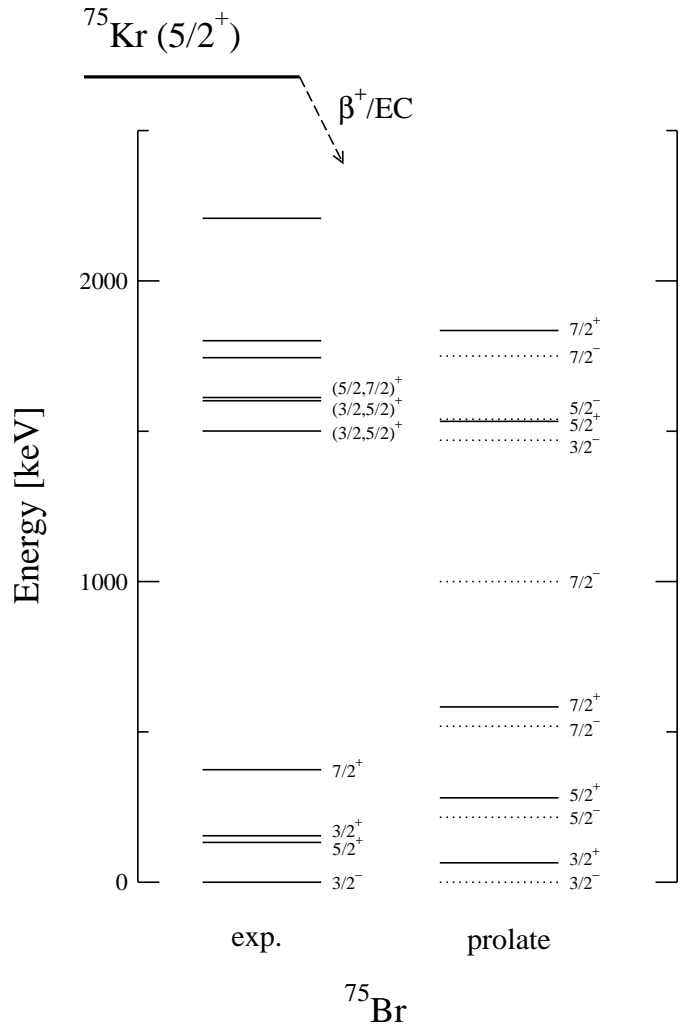,width=0.8\textwidth}
\caption{Same as in Fig. 11 for the decay of  $^{75}$Kr. }
\end{figure}
\vfill\eject
\begin{figure}[t]
\psfig{file=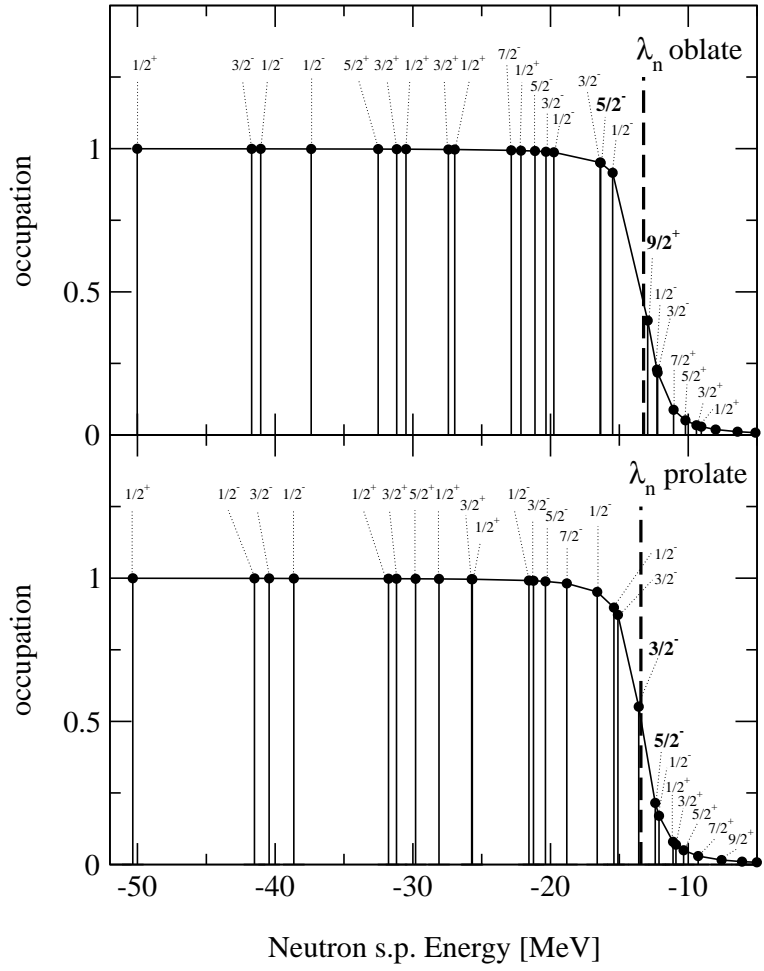,width=0.8\textwidth}
\caption{ Neutron single-particle energies and occupation probabilities 
of the intrinsic states $K^{\pi}$ calculated with the force SG2
for the two shapes, oblate and prolate,  that minimize the energy in $^{72}$Kr. 
$\lambda_n$ are the neutron Fermi energies.}
\end{figure}
\vfill\eject
\begin{figure}[t]
\psfig{file=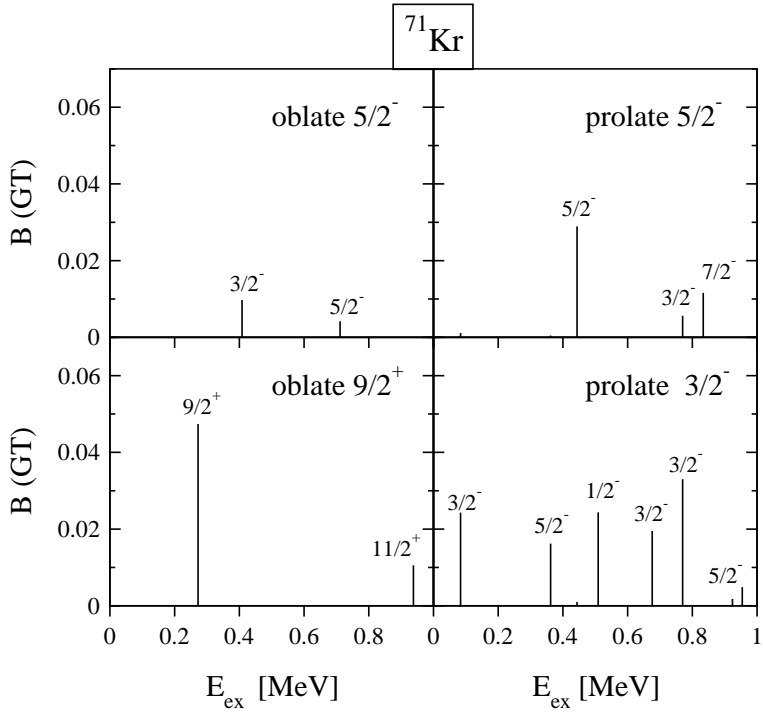,width=0.8\textwidth}
\caption{ Energy distribution of the Gamow-Teller strength in 
 $^{71}$Kr assuming different spins and parities of the ground state in
the parent nucleus. }
\end{figure}
\vfill\eject
\vfill\eject
\begin{figure}[t]
\psfig{file=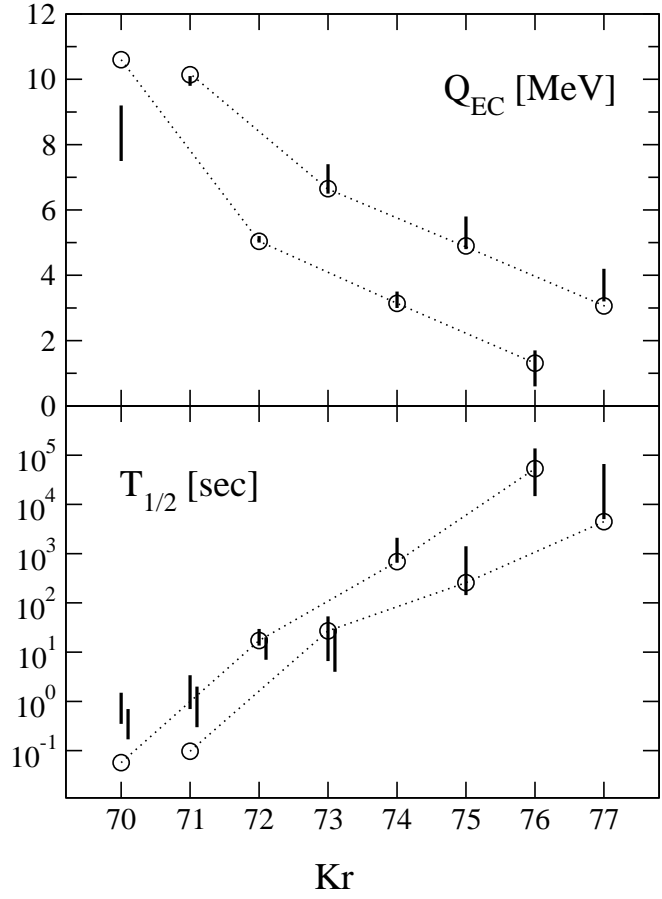,width=0.8\textwidth}
\caption{ Experimental $Q_{EC}$ values (top) and half-lives
(bottom) for the Kr isotopes are given by circles. The solid vertical lines
correspond to our theoretical QRPA calculations. The lengths of the vertical 
lines indicate the different results we obtain from using the forces SG2 and
Sk3 and the possible shapes in each isotope.}
\end{figure}
\vfill\eject
\end{document}